\documentclass[twocolumn,showpacs]{revtex4}
\usepackage[xdvi]{graphicx}%
\usepackage{dcolumn}
\usepackage{amsmath}
\makeatletter

\def\btt#1{\texttt{\@backslashchar#1}}%
\DeclareRobustCommand\bblash{\btt{\@backslashchar}}%
\makeatother

\newcommand{\bra}{\left\langle}
\newcommand{\ket}{\right\rangle}
\newcommand{\pder}[2]{\frac{\partial #1}{\partial  #2}}
\newcommand{\pdert}[2]{\frac{\partial^2 #1}{\partial  #2^2}}

\newcommand{\bv}[1]{{\boldsymbol #1}}

\newcommand{\tm}{\tau_{\rm m}}
\newcommand{\tM}{\tau_{\rm M}}
\newcommand{\tc}{\tau_{\rm c}}

\newcommand{\e}{{\rm e}}
\newcommand{\tsum}[1]{\sum_{#1=1}^{2}}

\begin{document}

\title{Anomalous time correlation in two-dimensional driven diffusive systems}
\author{Takenobu Nakamura}
\email{soushin@jiro.c.u-tokyo.ac.jp}
\author{Michio Otsuki}
\email{otsuki@jiro.c.u-tokyo.ac.jp}
\author{Shin-ichi Sasa}
\email{sasa@jiro.c.u-tokyo.ac.jp}

\affiliation
{Department of Pure and Applied Sciences,  
University of Tokyo, Komaba, Tokyo 153-8902, Japan}

\date{\today}

\begin{abstract} 
We study the time correlation function of a density field in 
two-dimensional driven diffusive systems within the framework 
of fluctuating hydrodynamics.
It is found that the time correlation exhibits power-law behavior in an 
intermediate time regime in the case 
that the fluctuation-dissipation relation is violated 
and that the power-law exponent depends on the extent 
of this violation. 
We obtain this result by employing a renormalization 
group method to treat a logarithmic divergence in time. 
\end{abstract}

\maketitle

\section{introduction}

The anomalous time correlation of hydrodynamic modes 
has been studied for a long period.
For an equilibrium fluid, it is understood that this anomaly arises
from nonlinear mode coupling effects \cite{pomeau}. 
By contrast, there is no systematic understanding of 
the time correlation in nonequilibrium steady states
(NESSs) far from  local equilibrium. In particular, it is not known how 
violation of the fluctuation-dissipation relation (FDR)
influences the time correlation.  


As the simplest example realizing a NESS far from local equilibrium,
we consider a two-dimensional driven diffusive system,
in which a fluctuating density field is driven locally by an external force. 
Such a system can be realized in laboratory experiments \cite{KTG}.
Perhaps the simplest model for a theoretical study of the long time 
behavior in the driven diffusive system is a stochastic differential 
equation consisting of terms representing a drift due to the external force, 
diffusion and random noise.


The time correlation function for such a stochastic model has been 
calculated by employing mode coupling theory \cite{BKS}. However,
the model analyzed in Ref. \cite{BKS} does not exhibit the long-range 
spatial correlation that is a generic feature of NESSs in driven 
diffusive systems of $d \ge 2$ dimensions.
The reason that long-range correlation does not appear in 
that model is that violation of the FDR is not taken into account.
Indeed, it is known that, in general, long-range correlation 
cannot exist when the FDR holds. By contrast, it has been found 
that the long-range correlation of driven diffusive systems 
can be described by a linear model with the violation of the FDR \cite{Grin}.

With the above considerations, in the present paper, we study a nonlinear 
model in which the FDR can be violated.
We demonstrate that as a result of 
this violation, the time correlation is qualitatively altered. 
Specifically, 
by employing a perturbative renormalization group (RG) method that 
treats a logarithmic divergence in time \cite{Illinois},
we obtain an expression for the time correlation function. 
From this expression, we find that 
power-law behavior appears in the time correlation 
if and only if the FDR is violated and that 
the power-law exponent depends on the extent of the violation.

\section{Model}

We consider the time evolution of a fluctuating density field
$\rho(\bv{x},t)$ in a two-dimensional space, under the 
influence of an external driving force in one direction, say the 
$x_1$ direction, where $\bv{x}=(x_1,x_2)$. 
Note that we study NESSs in the high temperature
regime, far from the critical point. We now describe the model 
we study. 
First, the conserved quantity $\rho$ obeys the continuity equation
\begin{equation}
\frac{\partial \rho(\bv{x},t)}{\partial t} 
+ \sum_{i=1}^2 \pder{J_i(\bv{x},t)}{x_i}=0.
\label{cont}
\end{equation}
We  assume that the $i$-th component of the density current, 
$J_i(\bv{x},t)$, is given by
\begin{equation}
J_i(\bv{x},t) 
= -  D_i \partial_i \rho(\bv{x},t)+ \delta_{i1}\bar J(\rho(\bv{x},t))
+ \xi_i(\bv{x},t).
\label{J}
\end{equation}
Here, the functional form of $\bar J$ is such that, 
with $\bar \rho$ the average density,
$\bar J(\bar \rho)$ is the average current along the $x_1$ direction 
in the steady state.
We then approximate $\bar J(\rho(\bv{x},t))$ in the form
\begin{equation}
\bar J(\rho(\bv{x},t))
\approx \bar J(\bar \rho) + c(\bar{\rho}) \delta \rho(\bv{x},t)
+ \lambda(\bar{\rho})(\delta \rho(\bv{x},t))^2,
\label{drift}
\end{equation}
where $\rho(\bv{x},t)=\bar \rho+\delta \rho(\bv{x},t)$.
The term $\xi_i(\bv{x},t)$ in (\ref{J}) represents a random current 
constituting  zero mean Gaussian white noise, with 
\begin{equation}
\bra \xi_i(\bv{x},t) \xi_j(\bv{x'},t')\ket 
=2B_i\delta_{ij}\delta(\bv{x}-\bv{x'})\delta(t-t').
\label{noise1}
\end{equation}
Note that because anisotropy in both the diffusion and noise intensity 
is expected to arise through effects of the external driving, 
the diffusion constant, $D_i$, 
and the noise intensity, $B_i$, are assumed to be anisotropic, in general.


Let us simplify the model given above. 
First, note that the first term on the right-hand side of (\ref{drift}) 
does not contribute to
the time evolution of the density, and 
the second term can be eliminated when we study density fluctuations 
in a frame moving with the velocity $c$ given in (\ref{drift}). 
To make this explicit, we define the density
$\phi(\bv{x},t) \equiv \delta \rho(\bv{x}+c\bv{e_1}t,t)$, 
where $\bv{e_1}$ is the unit vector in the $x_1$ direction. 
Furthermore, introducing the parameters $\chi$ and $\Delta$, 
we rewrite $B_1$ and $B_2$ as
\begin{equation}
B_i=D_i\chi(1-(-1)^i \Delta).
\label{delta:def}
\end{equation}
Thus, 
$\Delta$ corresponds to the extent of the violation of the FDR 
of the second kind \cite{kubo}.
Then, replacing $x_i$ by $\sqrt{D_i}x_i$,
$\phi$ by  
$\sqrt{\chi}(D_1D_2)^{1/4}\phi$
and  $\xi_i$ by $\sqrt{\chi D_i}(D_1D_2)^{-1/4}\xi_i$, 
we obtain the following dimensionless form of the 
equation for $\phi$:
\begin{equation}
\pder{\phi(\bv{x},t)}{t}
=\sum_{i=1}^{2}
 \left[\pdert{\phi(\bv{x},t)}{x_i}- \pder{\xi_i(\bv{x},t)}{x_i} \right]
-\bar \lambda \pder{\phi(\bv{x},t)^2}{x_1}.
\label{model}
\end{equation}
Here,
\begin{equation}
\bra \xi_i(\bv{x},t) \xi_j(\bv{x'},t')\ket 
=2\delta_{ij}(1-(-1)^i \Delta )\delta(\bv{x}-\bv{x'})\delta(t-t'),
\label{noise}
\end{equation}
and $\bar \lambda $ is a dimensionless constant given by
\begin{equation}
\bar \lambda=\lambda (D_1^3D_2)^{-1/4}\chi^{1/2}.
\end{equation}
The renormalization group flow of $(\bar \lambda, \Delta)$ 
for the model (\ref{model}) with (\ref{noise}) is  studied 
in Ref. \cite{ZS}. Also, the time correlation function has been  
calculated  in the  special cases that $\Delta=0$ (using the
mode coupling equation) \cite{BKS} and $\bar \lambda=0$ 
\cite{Grin}.  However, as far as we know, the time correlation function 
for the nonlinear model (\ref{model}) with the anisotropic noise 
intensity (\ref{noise}) has never been investigated.


In the analysis below,  employing a perturbative expansion with respect to 
$\bar \lambda$ and $\Delta$, we calculate the time correlation 
function $\hat{C}(\bv{k},t)$ defined by 
\begin{equation}
(2\pi)^2 \hat{C}(\bv{k},t)\delta(\bv{k}+\bv{k'})
= \bra \hat{\phi}(\bv{k},0)\hat{\phi}(\bv{k'},t)\ket. 
\label{simpleC}
\end{equation}
Here and below, for an arbitrary function $f(\bv{x},t)$, we define 
$\hat{f}(\bv{k},t)$ by
\begin{equation}
\hat{f}(\bv{k},t)\equiv\int d^2\bv{x}e^{-i\bv{k}\cdot\bv{x}}
f(\bv{x},t).
\label{Fourier}
\end{equation}
From the definition (\ref{simpleC}) and the symmetry of 
the steady state with respect to translation in time, 
the equality $\hat{C}(\bv{k},t)=\hat{C}(\bv{k},-t)$ holds.
Therefore, we consider only $\hat{C}(\bv{k},t)$ with $t \ge 0$.

\section{analysis}


First, we fix $\Delta$ and consider the expansion of $\hat\phi(\bv{k},t)$ 
in $\bar \lambda$:
\begin{align}
\hat \phi(\bv{k},t)=\hat \phi^{(0)}(\bv{k},t)
+\bar\lambda \hat \phi^{(1)}(\bv{k},t)
+\bar \lambda^2 \hat \phi^{(2)}(\bv{k},t)+\cdots.
\label{expansion1}
\end{align}
Substituting (\ref{expansion1}) into (\ref{model}) with (\ref{Fourier})
and extracting all terms proportional to $\bar \lambda^n$, we obtain a 
linear differential equation for $\hat\phi^{(n)}$ containing all lower order 
$\hat\phi^{(k)}$ and $\hat{\xi}_i(\bv{k},t)$. 
Solving these
differential equations under initial conditions set at $t=-\infty$,
we can iteratively derive  expressions for $\hat\phi^{(0)}$, 
$\hat\phi^{(1)}, \cdots $.  We then substitute 
these results into (\ref{simpleC}). In this way, 
the correlation function ${C}(\bv{k},t)$ is calculated in the form
\begin{equation}
\hat{C}(\bv{k},t)=\hat{C}^{(0)}(\bv{k},t)
+\bar \lambda\hat{C}^{(1)}(\bv{k},t)
+\bar \lambda^2\hat{C}^{(2)}(\bv{k},t)+\cdots. 
\end{equation}
It turns out that it is simplest to obtain the terms 
$\hat{C}^{(n)}(\bv{k},t)$ in the above expansion of $\hat{C}(\bv{k},t)$
by first deriving the terms $\tilde {C}^{(n)}(\bv{k},\omega)$
in the analogous expansion of $\tilde {C}(\bv{k},\omega)$,
the Fourier transform with respect to time of $\hat{C}(\bv{k},t)$,
and then taking the inverse Fourier transform of these.


The lowest-order contribution to $\hat{C}(\bv{k},t)$ 
can be easily calculated as 
\begin{equation}
\hat{C}^{(0)}(\bv{k},t)=
\left(1+ \Delta\frac{k_1^2-k_2^2}{|\bv{k}|^2}\right)\e^{-|\bv{k}|^2 t}.
\label{zeroth}
\end{equation}
Note that the spatial correlation function,
obtained through the Fourier transformation of $\hat{C}^{(0)}(\bv{k},0)$, 
exhibits  power-law decay of the type $1/r^2$, unless $\Delta=0$. 
This illustrates the long-range correlation of driven 
diffusive systems. To this order, we find that there is no interesting 
behavior of the time dependence of $\hat{C}^{(0)}(\bv{k},t)$, which merely 
exhibits an exponentially decaying form.


The next contribution to $\hat{C}(\bv{k},t)$ appears at second order 
in $\bar \lambda$. 
Through a straightforward calculation, we obtain 
\begin{widetext}
\begin{align}
\hat{C}^{(2)}(\bv{k},t)&=
-2\int_{-\infty}^{\infty}dt'\int\frac{d^2 \bv{k}'}{(2\pi)^2}
\frac{\sum_{i=1}^{2}(1-(-1)^i \Delta)k_i'^2}{|\bv{k}'|^2}
\e^{-|\bv{k}|^2|t-t'|-|\bv{k}'|^2|t'|-|\bv{k}-\bv{k}'|^2|t'|} \nonumber \\
&\left[
\frac{\sum_{j=1}^{2}(1-(-1)^j \Delta)k_j^2}
     {|\bv{k}|^2}k_1(k_1-k_1')\left((t-t')
\frac{t'}{|t'|}+ |t-t'|+\frac{1}{|\bv{k}|^2}\right)
\right.
-
\left.
\frac{1}{2}\frac{k_1^2}{|\bv{k}|^2}
\frac{\sum_{j=1}^{2}(1-(-1)^j \Delta)(k_j-k_j')^2}{|\bv{k}-\bv{k}'|^2}
\right].
\label{c2}
\end{align}
\end{widetext}
(In Sec. \ref{derivationc2}, we present a calculation method to
obtain this result.)
For this expression,  we first perform the integration over $|\bv{k'}|$ and 
then consider the $t'$ integration. Next, we carry out
the integration over the angle of $\bv{k'}$. However, this procedure 
is complicated by the fact that divergences appear in the
$t'$ integration. As one example, $\hat{C}^{(2)}(\bv{k},t)$ 
includes the term 
\begin{equation}
-\frac{1}{4\pi}k_1^2t\e^{-|\bv{k}|^2t}
\int_0^tdt'
\frac{1}{t'}\e^{|\bv{k}|^2t'/2} ,
\label{ex}
\end{equation}
where the contribution to the integral around $t'=0$ yields a
logarithmic divergence. Physically, this divergence
arises from the interaction between different modes during a very short 
time interval. However,  the model we study is assumed to 
be appropriate only for describing  phenomena over time scales longer 
than a certain scale $\tm$ in driven diffusive systems \cite{note:tm}, 
and this divergence should not exist in the case that we study a model 
that correctly describes the phenomena with  time scales shorter than 
$\tm$. However, here, instead of studying a model in which the 
microscopic details of behavior  on such short time scales are 
taken into account, we simply introduce a cut-off $\tm $; that is, 
the integration range of $t'$ in (\ref{c2}) 
is replaced by $[-\infty,-\tm] \cup [\tm,\infty]$.


With the cut-off introduced, the term (\ref{ex}) can be regularized as 
\begin{equation}
-\frac{1}{4\pi}k_1^2t\e^{-|\bv k|^2t}
\left[
\log\frac{t}{\tm}
+ \int^t_{\tm} dt' \frac{1}{t'}(\e^{|\bv{k}|^2 t'/2}-1)
\right].
\label{separation}
\end{equation}
Here, the first term exhibits a logarithmic divergence 
as $t/\tm \rightarrow \infty$, with fixed $|\bv{k}|^2 t$.
Following similar procedures, we can separate all singular 
terms from $\hat{C}^{(2)}(\bv{k},t)$, and in each case we obtain a term 
$\sim$ $\log t/\tm$.

Next, we expand $\hat{C}^{(2)}(\bv{k},t)$ in $\Delta$. 
Then,  to the first order, 
we obtain
\begin{align}
\hat{C}(\bv{k},t)&=\hat{C}^{(0)}(\bv{k},t) \nonumber \\
&\left[1- (c_0(\bv{k})\Delta+c_1(\Delta) k_1^2 t)
\bar \lambda^2 \log \frac{t}{\tm} \right]\nonumber \\
&+ \bar \lambda^2  \bar C^{(2)}(\bv{k},t) +o(\bar \lambda^2,\Delta),
\label{bare}
\end{align}
where $\bar C^{(2)}(\bv{k},t)$ represents the non-singular contribution
to $\hat{C}^{(2)}(\bv{k},t)$, and  $c_0(\bv{k})$ and $c_1(\Delta)$
are given by
\begin{eqnarray}
c_0(\bv{k})&=&
\frac{k_1^2}{8 \pi |\bv{k}|^2} \frac{k_1^2-3k_2^2}{|\bv{k}|^2}, 
\label{c0}\\
c_1(\Delta)&=& \frac{1}{8\pi}(2 - \Delta).
\label{c1}
\end{eqnarray}
(In Sec. \ref{derivationbare}, we present a more detailed explanation 
of the derivation.)
Note that the equal-time correlation $\hat{C}(\bv{k},0)$ 
must be obtained as 
$\lim_{\tm \to 0} \hat{C}(\bv{k},\tm)$, 
because the expression (\ref{bare}) 
is physically sound only for $t \ge \tm$.

The bare perturbation result (\ref{bare}) is reliable only for 
values of $t$ for which  $\log t/\tm$ is of  order  unity.
Now,  employing the RG method demonstrated in Ref. \cite{Illinois}, 
we derive a form of $\hat{C}(\bv{k},t)$ reliable even for $t \gg \tm$. 
First, we introduce a time scale $\tM$ which can be chosen arbitrarily 
and define a dimensionless parameter $\mu=\tM/\tm$. Then, using
\begin{equation}
\log \frac{t}{\tm}=\log  \frac{t}{\tM}+\log  \frac{\tM}{\tm},
\end{equation}
we  rewrite (\ref{bare}) as 
\begin{eqnarray}
\hat{C}(\bv{k},t)&=& Z(\mu ) \hat{C}^{(0)}(\bv{k},t) \nonumber \\
& \phantom{=}&
 \left[1- (c_0(\bv{k})\Delta +
         c_1(\Delta)  k_1^2 t)
\bar \lambda^2 \log \frac{t}{\tM}   \right]\nonumber \\
&+& \bar \lambda^2  \bar C^{(2)}(\bv{k},t) +o(\bar \lambda^2,\Delta),
\label{bare2}
\end{eqnarray}
where we have introduced  the renormalization constant 
$Z(\mu)$.
Here, the bare perturbation result (\ref{bare}) is 
equivalent to (\ref{bare2}) with
\begin{equation}
Z(\mu) =1- (c_0(\bv{k})\Delta 
         +c_1(\Delta)  k_1^2 t)
\bar \lambda^2 \log \mu +o(\bar \lambda^2,\Delta).
\label{z:bare}
\end{equation}
Now, we regard (\ref{z:bare}) as the bare perturbation
result for $Z(\mu )$ and  calculate  the improved 
perturbation result by using the fact that 
$\hat{C}(\bv{k}, t)$ does not depend on $\tM$. That is, differentiating (\ref{bare2})
with respect to $\tM$, we obtain the equation
\begin{equation}
\frac{ d \log Z(\mu)}{ d \log \mu}
+\left( c_0(\bv{k})\Delta +c_1(\Delta) k_1^2 t
 \right)\bar \lambda^2 +o(\bar \lambda^2,\Delta)=0,
\label{RG}
\end{equation}
which is referred to as the ``renormalization group equation''. 
Solving (\ref{RG}) under the condition 
\begin{equation}
Z(\mu = 0 ) = 1,
\end{equation}
we derive 
\begin{equation}
Z(\mu) =\mu^{-\left( c_0(\bv{k})\Delta +c_1(\Delta)  k_1^2 t\right)
 \bar \lambda^2 +o(\bar \lambda^2,\Delta)},
\label{Zflow}
\end{equation}
which provides the improved result of (\ref{z:bare}).  
Finally, substituting (\ref{Zflow}) into (\ref{bare2}) and setting 
$\tM=t$ (recall
that $\tM$ is arbitrary),  we obtain the expression
\begin{eqnarray}
  \hat{C}(\bv{k},t) &=& \hat{C}^{(0)}(\bv{k},t)
\left(\frac{t}{\tm}\right)^{
-\left( c_0(\bv{k})\Delta +c_1(\Delta)  k_1^2 t
\right)\bar \lambda^2 +o(\bar \lambda^2,\Delta)} \nonumber \\
&+&  \bar \lambda^2 \bar C^{(2)}(\bv{k},t)+o(\bar \lambda^2,\Delta),
\label{Rpert}
\end{eqnarray}
which may be reliable for all $ t \ge \tm$.

\section{Results and remarks}


From the expression (\ref{Rpert}), we have the following physically 
interesting results. 
First, we note that there is a crossover time $\tc(\bv{k})$ 
given by
\begin{equation}
\Delta |c_0(\bv{k})|\bar \lambda^2 
= |\bv {k}|^2 \tc(\bv{k}).
\end{equation}
We focus on the small wavenumber regime satisfying $\tc(\bv{k}) \gg \tm$. 
Then, for $t$ satisfying  $ t/\log(t/\tm) \ll \tc (\bv{k})$, 
the correlation of density fluctuations takes the power-law form
\begin{equation}
\hat{C}(\bv{k},t)\simeq 
\left(1+ \Delta\frac{k_1^2-k_2^2}{|\bv{k}|^2}\right)
\left(\frac{t}{\tm}\right)^{
-c_0(\bv{k})\Delta \bar \lambda^2 }.
\label{power}
\end{equation}
It is important to note that this power-law regime appears 
only when $\Delta \not=0$,
that is, only when the 
fluctuation-dissipation relation is violated [see (\ref{delta:def})]. 
We believe  that such power-law behavior can be observed in experiments. 

In addition to the above result, from (\ref{Rpert}), we find that
the decay rate of the correlation for $t$ satisfying
$ t/\log(t/\tm) \gg \tc(\bv{k})$ is expressed by
\begin{equation}
-\frac{1}{t}\log \hat{C}(\bv{k},t) \simeq 
|\bv{k}|^2+c_1(\Delta)k_1^2\bar \lambda^2 \log t.
\label{final}
\end{equation}
This shows that
the decay rate of the correlation increases slowly as a function 
of time.  Such enhancement of the decay rate exists even in the case 
$\Delta=0$. A similar result was obtained from  analysis 
of the mode coupling equation \cite{BKS}. 


The appearance of the singular term $\log t/\tm$ in the bare perturbation
result is the key to obtaining the power-law behavior of the correlation. 
A similar singular term was treated in Ref. \cite{Illinois} within 
the framework of the RG method to derive a solution representing  
anomalous diffusion for a deterministic nonlinear  diffusion equation. 
There are several related works 
\cite{JSS,CG} in which such a divergence is treated in a similar way. 

The RG analysis given here should not be confused with 
the method to study the RG flow of system parameters that
occurs with the change of the wavenumber scale. 
For the model under consideration, this type of the  RG flow 
of $(\bar \lambda, \Delta)$ is investigated in Ref. \cite{ZS}. 
With that method, for example, 
the relevancy of the parameters can be studied,
but explicit calculation of the time correlation is not possible.

In conclusion, we calculated the time correlation function (\ref{Rpert}) 
for the driven diffusive model (\ref{model}) with (\ref{noise}). 
The expression we obtained indicates that a power-law regime appears 
in the time correlation function if the FDR is violated. 
In addition to predicting this new type of physical phenomenon, 
our analysis 
provides an instructive example for the application of the perturbative RG 
method.

The authors thank K. Hayashi for her suggesting  the perturbative 
calculation of the time correlation function for the driven diffusive
system. 
This work was supported by a grant from the Ministry of 
Education, Science, Sports and Culture of Japan (No. 16540337).

\appendix
\section{derivation of (\ref{c2})}\label{derivationc2}
For an arbitrary function $f(\bv{x},t)$, 
we define $\tilde{f}(\bv{k},\omega)$ as
\begin{align}
\tilde{f}(\bv{k},\omega)
\equiv \int d^2\bv{x} dt
\e^{-i\omega t -i \bv{k}\cdot \bv{x}}
f(\bv{x},t) \label{zrepresentation}.
\end{align}
Then, the quantity $\tilde{C}(\bv{k},\omega)$ satisfies 
\begin{equation}
(2\pi)^3\delta(\bv{z}+\bv{z}')\tilde{C}(z)
=\bra
\tilde{\phi}(\bv{z})\tilde{\phi}(\bv{z}')
\ket,
\end{equation}
where $\bv{z}=(\bv{k},\omega)$. 
Here, the Fourier transformation of (\ref{model}) yields
\begin{equation}
\tilde{\phi}(\bv{z})=
G(\bv{z})\left[
-\tsum{i} ik_i\tilde{\xi}_i(\bv{z})
- \bar{\lambda} ik_1(\tilde{\phi}\circ\tilde{\phi})(\bv{z})
\right],\label{modelZ}
\end{equation}
with
\begin{equation}
G(\bv{z})\equiv
\frac{1}{i\omega + \tsum{i} k_i^2},
\end{equation}
where $(\tilde{f}\circ \tilde{g})(\bv{z})$ denotes the convolution of 
$\tilde{f}(\bv{z})$ and $\tilde{g}(\bv{z})$.
From (\ref{modelZ}), 
for $\tilde{\phi}^{(n)}(\bv{z}),(n=0,1,2,...)$, defined by
(\ref{expansion1}) and (\ref{zrepresentation}),
we obtain 
\begin{align}
\tilde{\phi}^{(0)}(\bv{z})
=&G(\bv{z})\left[
-\tsum{i} i k_i\tilde{\xi}_i(\bv{z})
\right],\\
\tilde{\phi}^{(1)}(\bv{z})
=&G(\bv{z})\left[
- i k_1 (\tilde{\phi}^{(0)}\circ\tilde{\phi}^{(0)})(\bv{z})
\right],\\
\tilde{\phi}^{(2)}(\bv{z})
=&G(\bv{z})\left[
- 2 i k_1 (\tilde{\phi}^{(0)}\circ\tilde{\phi}^{(1)})(\bv{z})
\right].
\end{align}
We expand $\tilde{C}(\bv{z})$ in the form 
\begin{multline}
\tilde{C}(\bv{z})
=\tilde{C}^{(0)}(\bv{z})
+\bar{\lambda}\tilde{C}^{(1)}(\bv{z})
+\bar{\lambda}^2\tilde{C}^{(2)}(\bv{z})
+... \ .
\label{correlationZ}
\end{multline}
The lowest order contribution of (\ref{correlationZ}) is calculated as
\begin{equation}
\tilde{C}^{(0)}(\bv{z})=
2|G(\bv{z})|^2\tsum{i}k_i^2(1+(-1)^{(i-1)}\Delta).
\end{equation}
Using the inverse Fourier transformation in $\omega$,
we obtain (\ref{zeroth}).
It can be easily checked $\tilde{C}^{(1)}(\bv{z})=0$, 
and $\tilde{C}^{(2)}(\bv{z})$ is expressed in the form
\begin{equation}
  \tilde{C}^{(2)}(\bv{z})
  =\tilde{C}^{(2)}_{\rm I}(\bv{z})
  +\tilde{C}^{(2)}_{\rm I\hspace{-.1em}I}(\bv{z})
    +\tilde{C}^{(2)}_{\rm I\hspace{-.1em}I\hspace{-.1em}I}(\bv{z}),
\label{sum}
\end{equation}
where 
\begin{widetext}
\begin{align}
\tilde{C}^{(2)}_{\rm I}(\bv{z})
=&8|G(\bv{z})|^2k_1^2 
\int d^3\bv{z}'
|G(\bv{z}-\bv{z}')|^2\tsum{i}(k_{i}-k_{i}')^2(1-(-1)^i \Delta)
|G(\bv{z}')|^2\tsum{j}k_{j}'^2(1-(-1)^j \Delta), \label{CI} \\
\tilde{C}^{(2)}_{\rm I\hspace{-.1em}I}(\bv{z})
=&32|G(\bv{z})|^4  \omega k_1\tsum{i}k_{i}^2(1-(-1)^i \Delta)
\int d^3\bv{z}'
|G(\bv{z}-\bv{z}')|^2 \tsum{j}(k_{j}-k_{j}')^2(1-(-1)^j \Delta)
| G(\bv{z}')|^2\omega'k_1', \label{CII} \\ 
\tilde{C}^{(2)}_{\rm I\hspace{-.1em}I\hspace{-.1em}I}(\bv{z})
=&-32|G(\bv{z})|^4 |\bv{k}|^2 k_1 
\tsum{i} k_{i}^2(1-(-1)^i \Delta)
\int d^3\bv{z}'
|G(\bv{z}-\bv{z}')|^2 \tsum{j}(k_{j}-k_{j}')^2(1-(-1)^j \Delta)
| G(\bv{z}')|^2 |\bv{k}'|^2 k_1'. \label{CIII}
\end{align}
\end{widetext}
Note that all the functions $\tilde{C}^{(2)}_{\alpha}(\bv{z})$, 
$(\alpha={\rm I,I\hspace{-.1em}I,I\hspace{-.1em}I\hspace{-.1em}I})$,
take the form
\begin{equation}
  \tilde{C}^{(2)}_{\alpha}(\bv{z})
  =\tilde{F}_{\alpha}(\bv{z})
  (\tilde{h}_{\alpha} \circ \tilde{\ell}_{\alpha})(\bv{z}),  
\end{equation}
where  $\tilde{F}_{\alpha}(\bv{z}),\tilde{h}_{\alpha}(\bv{z})$
and $\tilde{\ell}_{\alpha}(\bv{z})$ are
determined from (\ref{CI})-(\ref{CIII}).
Using this form, we can express $\hat{C}^{(2)}_{\alpha}(\bv{k},t)$ as
\begin{align}
\hat{C}^{(2)}_{\alpha}(\bv{k},t)
=&\int dt' \hat{F}_{\alpha}(\bv{k},t-t') \nonumber \\
&\int \frac{d^2 \bv{k}'}{(2\pi)^2}
\hat{h}_{\alpha}(\bv{k}-\bv{k}',t')\hat{\ell}_{\alpha}(\bv{k}',t') .
\label{mixconvolution}
\end{align}
Substituting this into (\ref{sum}),
we obtain (\ref{c2}).

\section{derivation of (\ref{bare})}\label{derivationbare}
We expand $\hat{C}^{(2)}(\bv{k},t)$ in the form 
\begin{equation}
\hat{C}^{(2)}(\bv{k},t)=
\hat{C}^{(2,0)}(\bv{k},t)
+ \Delta \hat{C}^{(2,1)}(\bv{k},t)+ o(\Delta) .
\end{equation}
Through a straightforward calculation, we obtain
\begin{equation}
\hat{C}^{(2,0)}(\bv{k},t)
= \frac{1}{4\pi}k_1^2
\left[\int_0^t dt'\e^{|\bv{k}|^2t'/2} 
-t\int_0^t dt'\frac{1}{t'}\e^{|\bv{k}|^2t'/2} \right].
\label{c2_0}
\end{equation}
In order to calculate $\hat{C}^{(2,1)}(\bv{k},t)$,
we extract terms proportional to $\Delta$ from (\ref{c2}).
The obtained expression becomes
\begin{widetext}
\begin{align}
\hat{C}^{(2,1)}(\bv{k},t)
=&-4\int_0^t dt'k_1(t-t')\e^{-|\bv{k}|^2(t-t')}
\int \frac{d^2\bv{k}'}{(2\pi)^2}
\left[\frac{k_1'^2-k_2'^2}{|\bv{k}'|^2}+\frac{k_1^2-k_2^2}{|\bv{k}|^2}\right]
(k_1-k_1')\e^{-[|\bv{k}-\bv{k}'|^2+|\bv{k}'|^2]t'}\nonumber \\ 
&+2\frac{k_1^2}{|\bv{k}|^2}\int dt'\e^{-|\bv{k}|^2|t-t'|}
\int \frac{d^2\bv{k}'}{(2\pi)^2}\frac{k_1'^2-k_2'^2}{|\bv{k}'|^2}
\e^{-[|\bv{k}-\bv{k}'|^2+|\bv{k}'|^2]|t'|}\nonumber \\
&-2\frac{k_1}{|\bv{k}|^2}\int dt'\e^{-|\bv{k}|^2|t-t'|}
\int \frac{d^2\bv{k}'}{(2\pi)^2}
\left[\frac{k_1'^2-k_2'^2}{|\bv{k'}|^2}+\frac{k_1^2-k_2^2}{|\bv{k}|^2}\right]
(k_1-k_1')\e^{-[|\bv{k}-\bv{k}'|^2+|\bv{k}'|^2]|t'|}.
\label{c2delta1}
\end{align}
\end{widetext}
We first evaluate the Gauss integrals in $|\bv{k}'|$ 
and perform the $t'$ integrals with picking up singular terms.
Then, $\hat{C}^{(2,1)}(\bv{k},t)$ is obtained as
\begin{align}
\hat{C}^{(2,1)}(\bv{k},t)
=-\frac{1}{8\pi}\e^{-|\bv{k}|^2t}k_1^2t
\left(\frac{k_1^2-3k_2^2}{|\bv{k}|^2}\right)\ln\frac{t}{\tm} \nonumber \\
-\frac{1}{8\pi}\e^{-|\bv{k}|^2t}\frac{k_1^2}{|\bv{k}|^2}
\left(\frac{k_1^2-3k_2^2}{|\bv{k}|^2}\right)\ln\frac{t}{\tm} \nonumber \\
+(\mbox{non-singular term}).\label{c2_1}
\end{align}
Combining (\ref{c2_0}) and (\ref{c2_1}) with (\ref{zeroth}),
we finally obtain (\ref{bare}) with (\ref{c0}) and (\ref{c1}).

\end{document}